\documentclass[aps,prd,twocolumn,showpacs,preprintnumbers,amsmath,amssymb,showpacs,showkeys]{revtex4}
\usepackage{graphicx}
\usepackage{dcolumn}
\usepackage{bm}
\usepackage{amssymb}
\usepackage{indentfirst}
\usepackage{epsfig}
\usepackage{epsf}
\usepackage{graphicx}
\usepackage{slashbox}

\newcommand{\eq}{\begin{eqnarray}}
\newcommand{\en}{\end{eqnarray}}
\newcommand{\bfb}{{\bf b}_{\perp}}
\newcommand{\bfk}{{\bf k}_{\perp}}

\newcommand{\ra}{\rangle}
\newcommand{\la}{\langle}

\begin{document}
 
\title{Generalized parton distributions in AdS/QCD}

\author{Alfredo Vega$^{1}$,
        Ivan Schmidt$^{1}$,
        Thomas Gutsche$^{2}$,
        Valery E. Lyubovitskij$^{2}$\footnote{On leave of absence
from Department of Physics, Tomsk State University, 634050 Tomsk, Russia}
\vspace*{1.2\baselineskip}}

\affiliation{$^{1}$Departamento de F\'\i sica y Centro Cient\'ifico y 
Tecnol\'ogico de Valpara\'iso,\\  
Universidad T\'ecnica Federico Santa Mar\'\i a,\\
Casilla 110-V, Valpara\'\i so, Chile \\
\vspace*{.2\baselineskip} \\
$^{2}$ Institut f\"ur Theoretische Physik,
Universit\"at T\"ubingen,\\
Kepler Center for Astro and Particle Physics, \\
Auf der Morgenstelle 14, D--72076 T\"ubingen, Germany
\vspace*{.8\baselineskip}}

\date{\today}

\begin{abstract}

The nucleon helicity-independent generalized parton distributions (GPDs) 
of quarks are calculated in the zero skewness case, in the framework of the AdS/QCD model.
The present approach is based on a matching procedure 
of sum rules relating the electromagnetic form factors to
GPDs and AdS modes. 

\end{abstract}



\preprint{USM-TH-277}

\maketitle

\section{Introduction}

One of the main goals in strong interaction theory is to understand 
how nucleons and other hadrons are build up from quarks and gluons. 
Studied in various scattering processes, the hadronic structure 
can be encoded in the so-called generalized parton distributions 
(GPDs)~\cite{Mueller:1998fv,Ji:1996nm,Radyushkin:1997ki,Radyushkin:1998rt}. 
In particular, at leading twist-2, there exist two kinds of 
helicity-independent GPDs of quarks in the nucleon, denoted as $H^q(x,\xi,t)$ 
and $E^q(x,\xi,t)$. Both quantities depend in general on three variables: 
the momentum transfer squared $t=q^2$, the light-cone momentum fraction $x$, 
and the skewness $\xi$. 

Due to their nonperturbative nature the GPDs cannot be directly 
calculated from Quantum Chromodynamics (QCD). There are essentially 
three ways to access the GPDs (for reviews see 
e.g.~\cite{Goeke:2001tz,Ji:2004gf}): extraction from the 
experimental measurement of hard processes, a direct 
calculation in the context of lattice QCD, and different phenomenological 
models and methods. The last procedure is based on a parametrization of 
the quark wave functions/GPDs using constraints imposed by 
sum rules~\cite{Ji:1996nm,Radyushkin:1997ki}, which relate
the parton distributions to nucleon electromagnetic form factors 
(some examples of this procedure can be found e.g. 
in~\cite{Diehl:2004cx,Guidal:2004nd,Selyugin:2009ic}). 
On the other hand, such sum rules can also be used in the other direction 
-- GPDs are extracted by calculating nucleon electromagnetic form 
factors in some approach. 

Following the last idea, here we show how to extract the quark GPDs 
of the nucleon in the framework of a holographical soft-wall 
model~\cite{Brodsky:2008pg,Abidin:2009hr}.  
In particular, we use the results of Abidin and Carlson 
for the nucleon form factors~\cite{Abidin:2009hr} in order to extract the GPDs 
using the light-front mapping -- the key ingredient of 
light-front holography~(LFH). This is an approach based 
on the correspondence of string theory in Anti-de Sitter~(AdS) 
space and conformal field theory~(CFT) 
in physical space-time~\cite{Maldacena:1997re}.  
LFH is further based 
on a mapping of string modes in the AdS fifth 
dimension to hadron light-front wave functions in physical space-time, 
as suggested and developed by Brodsky and 
de T\'eramond~\cite{Brodsky:2003px,Brodsky:2006uqa,Brodsky:2007hb,%
Brodsky:2008pg,Brodsky:2008pf} and extended 
in~\cite{Vega:2009zb,Vega:2010yq,Branz:2010ub}. 
In this paper we show how LFH can be used to get the nucleon GPDs 
in the context of the soft-wall model.  

From the beginning the AdS/CFT~\cite{Maldacena:1997re}  
correspondence has received 
considerable attention, which over time was expanded 
into several directions, one of which is the possibility
to address issues related to QCD phenomena. A particular
and easy way to consider AdS/CFT ideas applied to QCD is known as 
the bottom - up approach~\cite{Erlich:2005qh,DaRold:2005zs}, where one
tries to build models that reproduce some features of QCD in a dual 
5-dimensional space which contains gravity. This kind of models have been
successful in several QCD applications, among which are the following 
examples: hadronic scattering processes~\cite{Brodsky:2003px,%
Polchinski:2001tt,Janik:1999zk,Levin:2009vj}, 
hadronic spectra~\cite{Brodsky:2008pg,Branz:2010ub,Karch:2006pv,%
Vega:2008af,Vega:2010yr,Vega:2008te}, hadronic couplings and chiral 
symmetry breaking~\cite{Erlich:2005qh,DaRold:2005zs,Grigoryan:2007my,%
Colangelo:2008us,Vega:2010ne}, 
quark potentials~\cite{BoschiFilho:2005mw,Andreev:2006ct,Jugeau:2008ds}, 
etc. 

In this paper we perform a matching of the
nucleon electromagnetic form factors 
considering two approaches for them: we use sum rules derived in
QCD~\cite{Ji:1996nm,Radyushkin:1997ki}, which contain
GPDs for valence quarks, and we consider 
an expression obtained in the AdS/QCD soft-wall model~\cite{Abidin:2009hr}.  
As a result of the matching we obtain expressions for 
the nonforward parton densities~\cite{Radyushkin:1998rt}  
$H_{v}^{q}(x,t) = H^q(x,0,t) + H^q(-x,0,t)$ and 
$E_{v}^{q}(x,t) = E^q(x,0,t) + E^q(-x,0,t)$ -- 
flavor combinations of the GPDs (or valence GPDs), 
using information from the AdS side. 
The procedure proposed here is similar to the one used in LFH,  
which allows to obtain a light front wave function related to 
the AdS modes associated with mesons~\cite{Brodsky:2003px,Brodsky:2006uqa,%
Brodsky:2007hb,Brodsky:2008pg,Brodsky:2008pf}. 
Contrary to the LFH approach, here 
the holographical coordinate is not considered as 
a parton distance in hadrons, so we do not need to propose 
a modification in the AdS/CFT dictionary. 
Also we look at several impact space properties of the nucleons: 
impact parameter dependent GPDs, parton charge densities in the 
transverse impact space, transverse widths and root mean 
square (rms) radii~\cite{Burkardt:2000za,Ralston:2001xs,Diehl:2002he,%
Belitsky:2002ep,Ji:2003ak,Diehl:2004cx,Miller:2007uy}.  
 
The nucleon electromagnetic form factors $F_1^N$ and $F_2^N$ 
($N=p, n$ correspond to proton and neutron) are conventionally 
defined by the matrix element of the electromagnetic current as 
\eq 
\label{CorrienteEM}
\langle p' | J^{\mu}(0) | p \rangle = 
\bar{u}(p') [ \gamma^{\mu} F_{1}^N(t) 
+ \frac{i \sigma^{\mu \nu}}{2 m_N}  \, q_\nu 
F_{2}^N(t)] u(p),
\en 
where $q = p' - p$ is the momentum transfer; 
$m_N$ is the nucleon mass; and $F_1^N$ and $F_2^N$ are the Dirac and Pauli form factors, which 
are normalized to the electric charge $e_N$ and anomalous 
magnetic moment $k_N$ of the corresponding nucleon: 
$F_1^N(0)=e_N$ and $F_2^N(0)=k_N$.  

The sum rules relating the electromagnetic form factors and the GPDs read 
as~\cite{Ji:1996nm,Radyushkin:1997ki,Radyushkin:1998rt}  
\eq 
F_{1}^{p}(t) &=& \int_{0}^{1} dx \, \biggl( \frac{2}{3}H_{v}^{u}(x,t) 
                                  - \frac{1}{3}H_{v}^{d}(x,t)\biggr)\,, 
\nonumber\\
F_{1}^{n}(t) &=& \int_{0}^{1} dx \, \biggl( \frac{2}{3}H_{v}^{d}(x,t) 
                                  - \frac{1}{3}H_{v}^{u}(x,t)\biggr)\,, 
\nonumber\\  
&&\\                                  
F_{2}^{p}(t) &=& \int_{0}^{1} dx \, \biggl( \frac{2}{3}E_{v}^{u} (x,t) 
                                  - \frac{1}{3}E_{v}^{d} (x,t)\biggr)\,,
\nonumber\\ 
F_{2}^{n}(t) &=& \int_{0}^{1} dx \, \biggl( \frac{2}{3}E_{v}^{d} (x,t) 
\nonumber                         - \frac{1}{3}E_{v}^{u} (x,t)\biggr)\,, 
\en 
Here we restrict our analysis to the contribution of the $u$ and $d$ 
quarks and antiquarks, while the presence of the heavier strange and charm 
quark constituents is not considered. 

\section{GPDs in AdS/QCD} 

\subsection{Electromagnetic nucleon form factors}

In order to derive the GPDs in AdS/QCD we outline the relevant 
results obtained by Abidin and Carlson~\cite{Abidin:2009hr} 
for the nucleon form factors using an 
AdS/QCD model. It is based on soft-wall breaking of conformal invariance 
by introducing a quadratic dilaton field $\Phi (z) = \kappa^{2} z^{2}$ 
in the action (in the overall exponential and in 
the mass term)~\cite{Abidin:2009hr}. Such a procedure leads to 
Regge-like mass spectra in the baryonic sector. Note that a similar 
AdS/QCD approach for baryons was developed by Brodsky and 
de T\'eramond in~\cite{Brodsky:2008pg}. One should stress that 
introduction of the dilaton field in both approaches is based on the 
idea of getting the simplest analytical solution of the equations of 
motion of the string mode. Further corrections like higher powers 
in the holographic coordinate can be included, although they do not 
change the physics significantly. The AdS metric is specified as 
\begin{equation}
\label{Metrica}
 ds^{2} = g_{MN} dx^M dx^N = 
\frac{1}{z^{2}} (\eta_{\mu \nu} dx^{\mu} dx^{\nu} - dz^{2}),
\end{equation}
where $\mu, \nu = 0, 1, 2, 3$; $\eta_{\mu \nu} = {\rm diag}(1,-1,-1,-1)$ 
is the Minkowski metric tensor 
and $z$ is the holographical coordinate running from zero to $\infty$. 

The relevant terms in the AdS/QCD action, which generate the nucleon 
form factors, are~\cite{Abidin:2009hr}: 
\eq 
S &=& \int d^4x \, dz \, \sqrt{g} \, e^{-\Phi(z)} \, \Big( \bar\Psi 
\, e_A^M \, \Gamma^A \, V_M \, \Psi \, \nonumber\\
&+& \frac{i}{2} \, \eta_{S,V} \, \bar\Psi \, e_A^M \, e_B^N \, 
[\Gamma^A, \Gamma^B] \, F^{(S,V)}_{MN} \, \Psi \, \Big) \,, 
\en 
where the basic blocks of the AdS/QCD model are defined 
as~\cite{Abidin:2009hr}: 
$g = |{\rm det} \, g_{MN}|$; $\Psi$ and $V_M$ are the 5D Dirac and vector 
fields dual to the nucleon and electromagnetic fields, respectively; 
$F_{MN} = \partial_M V_N - \partial_N V_M$;  
$\Gamma^A = (\gamma^\mu, - i \gamma^5)$; 
$e_A^M = z \delta_A^M$ is the inverse vielbein; and $\eta_{S,V}$ are the couplings constrained by the  
anomalous magnetic moment of the nucleon: 
$\eta_p = (\eta_S + \eta_V)/2 = \kappa \,k_p/(2m_N\sqrt{2})$ 
and   
$\eta_n = (\eta_S - \eta_V)/2 = \kappa \,k_n/(2m_N\sqrt{2})$.  
Here the indices $S,V$ denote isoscalar and isovector 
contributions to the electromagnetic form factors. 

Finally, the results for the nucleon form factors in AdS/QCD 
are given by~\cite{Abidin:2009hr}: 
\eq 
F_{1}^p(Q^2) &=& C_{1}(Q^2) + \eta_{p} C_{2}(Q^2)\,,\nonumber\\
F_{2}^p(Q^2) &=& \eta_{p} C_{3}(Q^2)\,,\nonumber\\
& &\\
F_{1}^n(Q^2) &=& \eta_{n} C_{2}(Q^2)\,,\nonumber\\
F_{2}^n(Q^2) &=& \eta_{n} C_{3}(Q^2), \nonumber 
\en 
where $Q^{2} = - t$ and 
$C_i(Q^2)$ are the structure integrals: 
\eq 
C_{1}(Q^2) &=& \int dz e^{-\Phi} 
\frac{V(Q,z)}{2 z^{3}} (\psi_{L}^{2}(z) + \psi_{R}^{2}(z))\,, 
\nonumber\\ 
C_{2}(Q^2) &=& \int dz e^{-\Phi} \frac{\partial_{z} V(Q,z)}{2 z^{2}} 
(\psi_{L}^{2}(z) - \psi_{R}^{2}(z))\,,  \label{Ci}\\ 
C_{3}(Q^2) &=& \int dz e^{-\Phi} 
\frac{2 m_{N} V(Q,z)}{2 z^{2}} \psi_{L}(z) \psi_{R}(z)\,. 
\nonumber 
\en 
$\psi_{L}(z)$ and $\psi_{R}(z)$ are the Kaluza-Klein modes  
(normalizable wave functions), which are dual to left- and right-handed 
nucleon fields:  
\eq \label{PsiLR}
\psi_L(z) = \kappa^3 z^4\,, \quad 
\psi_R(z) = \kappa^2 z^3 \sqrt{2} 
\en 
and 
\eq\label{V}
V (Q,z) = \Gamma(1 + \frac{Q^{2}}{4 \kappa^{2}}) 
U(\frac{Q^{2}}{4 \kappa^{2}}, 0, \kappa^2 z^2)
\en 
is the bulk-to-boundary propagator of the vector field 
in the axial gauge. 
Note that expressions for the nucleon form factors in AdS/QCD can be 
presented in an analytical form after integration over the variable $z$.  
In particular, the $C_i$ functions, defining the Dirac and Pauli  
factors, are given by:   
\eq
C_1(Q^2) &=& \frac{a+6}{(a+1)(a+2)(a+3)} \,, \nonumber\\
C_2(Q^2) &=& \frac{2a (2a-1)}{(a+1)(a+2)(a+3)(a+4)} \,, \\
C_3(Q^2) &=& \frac{12 m_N \sqrt{2}}{\kappa} \,  
\frac{1}{(a+1)(a+2)(a+3)} \,, \nonumber 
\en 
where $a = Q^2/(4\kappa^2)$. Note that we obtain the correct 
scaling behavior of the nucleon form factors at large $Q^2$, 
$F_1^{p,n} \sim 1/Q^4$ and $F_2^{p,n} \sim 1/Q^6$~\cite{Abidin:2009hr}. 
Also we get reasonable agreement for the 
slopes of the nucleon form factors with data. In particular, in terms 
of the nucleon magnetic moments $\mu_p = 1 + k_p$ and $\mu_n = k_n$ the 
charge $(r_E^p, r_E^n)$ and magnetic $(r_M^p, r_M^n)$ 
radii are written as
\eq 
\la r^2_E \ra^p &=& \frac{147}{64 \kappa^2} 
\biggl( 1 + \frac{13}{147} \mu_p \biggr) \,, \nonumber\\ 
\la r^2_E \ra^n &=& \frac{13}{64 \kappa^2} \mu_n\,, \\
\la r^2_M \ra^p &=& \frac{177}{64 \kappa^2} 
\biggl( 1 - \frac{17}{177 \mu_p} \biggr) \,, \nonumber\\ 
\la r^2_M \ra^n &=& \frac{177}{64 \kappa^2} \,. \nonumber
\en 
Notice that in the context of AdS/QCD 
charge radii have been discussed before in~\cite{Abidin:2009hr}. 
Our numerical results for the slopes compared rather well with data: 
\eq 
\la r^2_E \ra^p &=& 0.910 \ {\rm fm}^2 \ {\rm (our)}\,, 
\quad 
0.766 \ {\rm fm}^2 \ {\rm (data)}\,, \nonumber\\
\la r^2_E \ra^n &=& - 0.123 \ {\rm fm}^2 \ {\rm (our)}\,, 
\quad  
- 0.116 \ {\rm fm}^2 \ {\rm (data)}\,, \nonumber\\
\la r^2_M \ra^p &=& 0.849 \ {\rm fm}^2 \ {\rm (our)}\,, 
\quad 
0.731 \ {\rm fm}^2 \ {\rm (data)}\,, \\
\la r^2_M \ra^n &=& 0.879 \ {\rm fm}^2 \ {\rm (our)}\,, 
\quad 
0.762 \ {\rm fm}^2 \ {\rm (data)}\,. \nonumber
\en

\subsection{Nucleon GPDs in momentum space} 

Expressions for the GPDs in terms of the AdS modes can be obtained 
using the procedure of light-front mapping suggested by Brodsky and de
T\'eramond~\cite{Brodsky:2007hb}.
In the present case this procedure is based 
on the use of the integral representation for the 
bulk-to-boundary propagator introduced by 
Grigoryan and Radyushkin~\cite{Grigoryan:2007my}: 
\eq 
\label{VInt}
V (Q,z) = \kappa^{2} z^{2} \int_{0}^{1} \frac{dx}{(1-x)^{2}} 
x^{\frac{Q^{2}}{4 \kappa^{2}}} 
e^{- \displaystyle{\frac{\kappa^2 z^2 x}{1-x} }}\,,   
\en 
where the variable $x$ is equivalent to the light-cone momentum 
fraction~\cite{Brodsky:2007hb}.  
Matching the respective expressions for the nucleon form factors results 
(after performing the integration over the holographic coordinate $z$) 
in the nonforward parton densities of the nucleon as: 
\eq 
H_{v}^{q}(x,Q^2) &=& q(x) \, x^a \,, 
\label{HqAdS}\\ 
E_{v}^{q}(x,Q^2) &=& e^q(x) \, x^a \,. 
\label{EqAdS}  
\en 

\begin{figure}
  \begin{tabular}{c}
    \includegraphics[width=3.0 in]{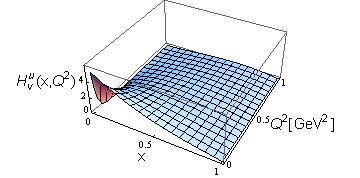}\\
    \includegraphics[width=3.0 in]{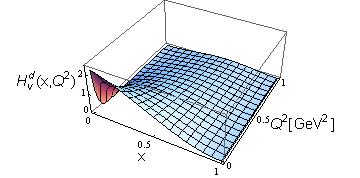}
  \end{tabular}
\caption{$H_{v}^{q} (x,Q^2)$ in the holographical model.}

  \begin{tabular}{c}
\vspace*{.25cm} 
    \includegraphics[width=3.0 in]{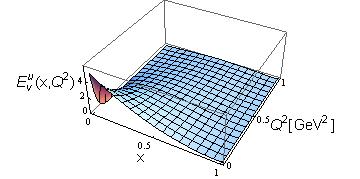}\\
    \includegraphics[width=3.0 in]{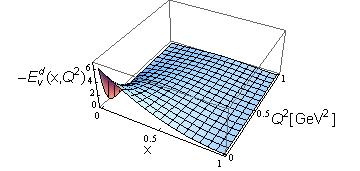}
  \end{tabular}
\caption{$E_{v}^{q} (x,Q^2)$ in the holographical model.}
\end{figure}

Here $q(x)$ and $e^q(x)$ are distribution functions given by: 
\eq 
q(x)   = \alpha^q \gamma_{1}(x) + \beta^q \gamma_{2}(x)\,, \quad 
e^q(x) = \beta^q \gamma_{3}(x)\,, 
\en 
where the flavor couplings $\alpha^q, \beta^q$ 
and functions $\gamma_i(x)$ are written as
\eq 
\alpha^u = 2\,, \ \alpha^d = 1\,, \ 
\beta^u = 2 \eta_{p} + \eta_{n} \,, \ 
\beta^d = \eta_{p} + 2 \eta_{n} \,  
\en 
and 
\eq
\gamma_{1}(x) &=& 
\frac{1}{2} (5 - 8x + 3x^{2})\,, \nonumber\\
\gamma_{2}(x) &=& 1 - 10x + 21x^{2} - 12x^{3} \,, 
\label{gamma} \\
\gamma_{3}(x) &=& 
\frac{6 m_N \sqrt{2}}{\kappa} (1 - x)^{2} \,. \nonumber
\en 
Eqs.~(\ref{HqAdS})-(\ref{gamma}), which display the nonforward parton 
densities of the nucleon, are the main result of this matching procedure. 
Notice that these functions have an exponential form, which is typical 
when choosing an ansatz for these functions. The distribution functions 
are also consistent with a linear Regge 
behavior at small $x$~\cite{Goeke:2001tz,Guidal:2004nd}. 
In Figs.1 and 2 we show the nonforward parton distributions 
$H_{v}^{q}$ and $E_{v}^{q}$ 
for nucleons, obtained from the expressions deduced on the AdS side, 
according to the holographical model considered
in~\cite{Abidin:2009hr}.

The parameters involved are the same as used in~\cite{Abidin:2009hr}, 
i.e. $\kappa = 350$ MeV, $\eta_{p} = 0.224$,
$\eta_{n} = -0.239$, which were
fixed in order to reproduce the mass $m_N = 2\kappa \sqrt{2}$ and 
the anomalous magnetic moments of the nucleon 
$k_p = \mu_p - 1 = 1.791$ and $k_n = \mu_n = - 1.913$. 

\begin{center}
\begin{figure*}[ht]
  \begin{tabular}{c c}
    \includegraphics[width=3.0 in]{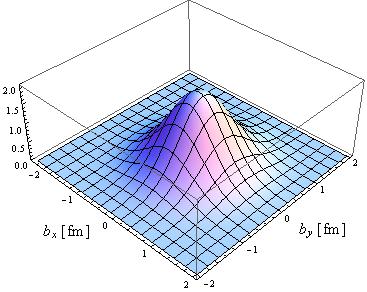} \hspace*{1cm}
    \includegraphics[width=3.0 in]{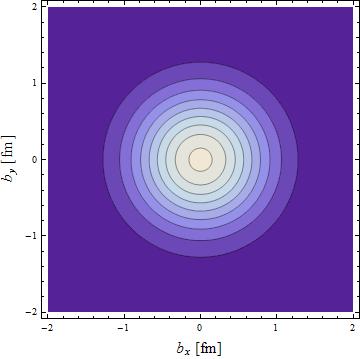} \\
    \includegraphics[width=3.0 in]{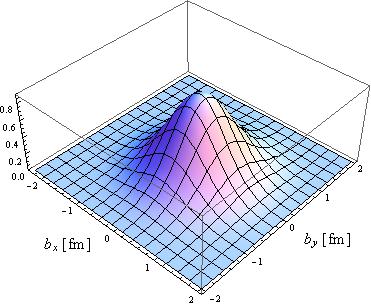} \hspace*{1cm}
    \includegraphics[width=3.0 in]{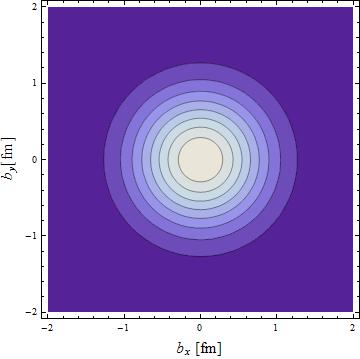}
  \end{tabular}
\caption{Plots for $q(x,\bfb)$. The upper panels correspond to $u(x,\bfb)$ and 
the lower to $d(x,\bfb)$. Both cases are taken for $x = 0.1$.}
\end{figure*}
\end{center}

For completeness we also analyze the moments of the valence GPDs 
$H^q_v(x,Q^2)$ and $E^q_v(x,Q^2)$~\cite{Diehl:2004cx}: 
\eq\label{GPD_moments} 
h^q_n(Q^2) &=& \int\limits_{0}^1 dx x^{n-1} H^q_v(x,Q^2)\,, \\ 
e^q_n(Q^2) &=& \int\limits_{0}^1 dx x^{n-1} E^q_v(x,Q^2)\,.  
\en 
Integration over $x$ results in: 
\eq 
h^q_n(Q^2) &=& \alpha^q \, \frac{n+a+5}{(n+a) \ldots (n+a+2)} \nonumber\\
&+& 4 \beta^q \, \frac{(n+a-1)(n+a-3/2)}{(n+a) \ldots (n+a+3)}\,, \\
e^q_n(Q^2) &=& \frac{12 \beta^q m_N \sqrt{2}}{\kappa} \, 
\frac{1}{(n+a) \ldots (n+a+2)} \,. \nonumber 
\en
It can be useful to compare our predictions for the first moments 
$h^q_1(Q^2)$ and $e^q_1(Q^2)$ with the available lattice results of 
Ref.~\cite{Hagler:2004er}. These lattice predictions have been 
approximated by the dipole form formulas: 
\eq 
h^q_1(Q^2) &=& \frac{h^q_1(0)}{(1+Q^2/M_h^2)^2}\,, \nonumber\\ 
e^q_1(Q^2) &=& \frac{e^q_1(0)}{(1+Q^2/M_e^2)^2}\,,  
\en 
where $M_h = 1.47 \pm 0.03$ GeV and $M_e = 1.16 \pm 0.02$ GeV are 
the dipole mass parameters. Then the slopes 
of the lattice form factors $h^q_1(Q^2)$ and $e^q_1(Q^2)$ are: 
\eq 
\la r_h^2 \ra &=& - 6 \frac{d\log h^q_1(Q^2)}{dQ^2}\bigg|_{Q^2 = 0} 
\simeq 0.216 \ {\rm fm}^2\,, \nonumber\\ 
\la r_e^2 \ra &=& - 6 \frac{d\log e^q_1(Q^2)}{dQ^2}\bigg|_{Q^2 = 0} 
\simeq 0.347 \ {\rm fm}^2\,. 
\en 
In our approach the slopes $\la r_h^2 \ra$ and $\la r_e^2 \ra$ are given 
by 
\eq 
\la r_h^2 \ra &=& \frac{5}{2 \kappa^2} 
\biggl( 1 + \frac{\beta^q}{20 \alpha^q} \biggr) \,, \nonumber\\
\la r_e^2 \ra &=& \frac{11}{4 \kappa^2} \,. 
\en 
As for the radii of nucleon electromagnetic form factors, these slopes 
are proportional to the $1/\kappa^2$ and are therefore well 
constrained. Since 
our predictions for the nucleon electromagnetic radii are in agreement 
with data for a value of $\kappa = 350$ MeV, fixed from the nucleon mass, our predictions for the slopes of the first 
moments of the nucleon GPDs should also be consistent with data.  
In particular, we get:  
$\la r_h^2 \ra = 0.800$ fm$^2$ for the $u-$quark and 
0.785 fm$^2$ for the $d-$quark; 
$\la r_e^2 \ra = 0.874$ fm$^2$ (independent on the quark flavor). 
Our results for the slopes $\la r_h^2 \ra$ 
and $\la r_e^2 \ra$ are larger in comparison to lattice predictions. 
Obviously, more accurate lattice results 
at the physical value of the pion mass are necessary. 

\subsection{Nucleon GPDs in impact space}

Another interesting aspect to consider is the nucleon GPDs 
in impact space. As shown by Burkardt~\cite{Burkardt:2000za}, 
the GPDs in momentum space are related to the impact parameter 
dependent parton distributions by a Fourier transform. 
GPDs in impact space give access to 
the distribution of partons in the transverse plane, which is 
quite important for understanding the nucleon structure.

\begin{center}
\begin{figure*}[t]
  \begin{tabular}{c c}
    \includegraphics[width=3.0 in]{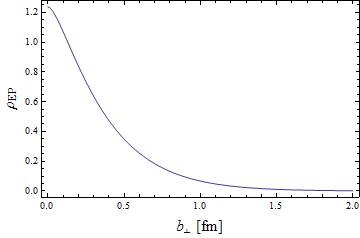}
    \includegraphics[width=3.0 in]{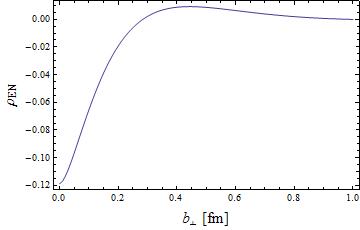} \\
    \includegraphics[width=3.0 in]{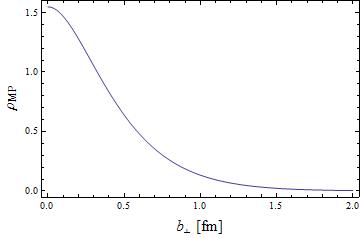}
    \includegraphics[width=3.0 in]{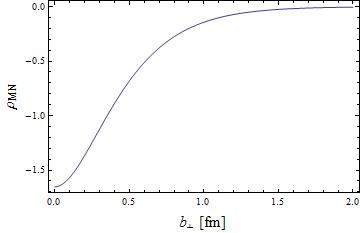}
  \end{tabular}
\caption{Parton charge $\rho_{EN} \equiv \rho_E^N(\bfb)$ 
and magnetization $\rho_{MN} \equiv \rho_M^N(\bfb)$  
densities in the transverse impact space for proton $(N=p)$ and
neutron $(N=n)$.}
\end{figure*}
\end{center}

Following Refs.~\cite{Burkardt:2000za} 
and~\cite{Diehl:2004cx,Miller:2007uy} 
we define the following set of nucleon quantities in impact space: 
i) the nucleon GPDs in impact space 
\eq 
q(x,\bfb) &=& \int\frac{d^2\bfk}{(2\pi)^2} H_q(x,\bfk^2) 
e^{-i\bfb\bfk}\,, \nonumber\\
e^q(x,\bfb) &=& \int\frac{d^2\bfk}{(2\pi)^2} E_q(x,\bfk^2) 
e^{-i\bfb\bfk}\,, 
\en 
ii) parton charge $\rho_E^N(\bfb)$ 
and magnetization $\rho_M^N(\bfb)$ densities in 
transverse impact space 
\eq 
\rho_E^N(\bfb) &=& \sum\limits_{q} e_q^N \int\limits_0^1 dx q(x,\bfb) \,,
\nonumber\\
\rho_M^N(\bfb) &=& \sum\limits_{q} e_q^N \int\limits_0^1 dx e^q(x,\bfb) \,,
\en 
where $e_u^p = e_d^n = 2/3$ and $e_u^n = e_d^p = - 1/3$, 

iii) transverse width of the impact parameter dependent GPD 
$q(x,\bfb)$ 
\eq 
\la R_\perp^2(x) \ra_q &=& 
\frac{\int d^2\bfb \bfb^2 q(x,\bfb)}
{\int d^2\bfb q(x,\bfb)} \nonumber\\
&=& 
- 4 \frac{\partial \log H_v^q(x,Q^2)}{\partial Q^2}\bigg|_{Q^2=0} \,, 
\en 
iv) transverse rms radius 
\eq 
\la R_\perp^2 \ra_q = 
\frac{\int d^2\bfb \bfb^2 \int\limits_0^1 dx q(x,\bfb)}
{\int d^2\bfb  \int\limits_0^1 dx q(x,\bfb)} \,. 
\en 
Notice that the GPDs in impact space can be derived directly from 
the nucleon form factors using the procedure of light-front 
mapping and the bulk-to-boundary propagator in impact 
space $V(\bfb,z)$. The latter is related to $V(\bfk,z)$ via 
the Fourier transform: 
\eq 
V(\bfb,z) &=& \int\frac{d^2\bfk}{(2\pi)^2} V(\bfk,z) 
e^{-i\bfb\bfk} \nonumber\\
&=& \frac{\kappa^4 z^2}{\pi} \int\limits_0^1 dx 
\displaystyle{\frac{e^{\displaystyle{- \frac{\kappa^2z^2 x}{1-x} 
- \frac{\bfb^2\kappa^2}{\log(1/x)}}}}{(1-x)^2 \log(1/x)}} \,. 
\en 
Soft-wall AdS/QCD gives the following predictions for 
the impact space properties of nucleons: 
\eq 
q(x,\bfb) &=& q(x) \frac{\kappa^2}{\pi\log(1/x)} 
e^{- \frac{\bfb^2\kappa^2}{\log(1/x)}} \,, \nonumber\\
e^q(x,\bfb) &=& e^q(x) \frac{\kappa^2}{\pi\log(1/x)} 
e^{- \frac{\bfb^2\kappa^2}{\log(1/x)}} \,, \nonumber\\ 
\rho_E^N(\bfb) &=& \frac{\kappa^2}{\pi}
\sum\limits_{q} e_q^N \int\limits_0^1 \frac{dx}{\log(1/x)}
q(x) e^{- \frac{\bfb^2\kappa^2}{\log(1/x)}} \,, \nonumber\\
\rho_M^N(\bfb) &=& \frac{\kappa^2}{\pi}
\sum\limits_{q} e_q^N \int\limits_0^1 \frac{dx}{\log(1/x)}
e^q(x) e^{- \frac{\bfb^2\kappa^2}{\log(1/x)}} \,, \nonumber\\ 
\la R_\perp^2(x) \ra_q &=& \frac{\log(1/x)}{\kappa^2} \,, \nonumber\\
\la R_\perp^2\ra_q &=& \frac{1}{\kappa^2}
\biggl( \frac{5}{3} + \frac{\beta^q}{12 \alpha^q} \biggr) \,. 
\en  
Fig.3 shows some examples of $q(x,\bfb)$ and in Fig.4 
we plot $\rho_E^N(\bfb)$ and $\rho_M^N(\bfb)$ for proton and neutron. 
For the transverse rms radius of $u-$ and $d-$quark GPDs we get 
similar values: 
\eq 
\la R_\perp^2\ra_u = 0.527 \ {\rm fm}^2\,, \quad 
\la R_\perp^2\ra_d = 0.524 \ {\rm fm}^2\,.  
\en 
One should stress that the obtained nucleon GPDs both in momentum and 
impact spaces correspond to the so-called ``Gaussian ansatz'' and 
are consistent with general predictions for their asymptotic  
behavior for $x \to 0$ or $x \to 1$ and $Q^2 \to 0$ or 
$Q^2 \to \infty$~\cite{Burkardt:2000za,Guidal:2004nd}. 

\section{Conclusions} 

We determined the nucleon GPDs both in momentum and impact 
space using ideas of AdS/QCD, LFH and 
sum rules relating electromagnetic form factors to the 
GPD functions $H_{v}^{q}(x,Q^2)$ and $E_{v}^{q}(x,Q^2)$.  
The procedure used is similar to the one considered in some
applications of LFH, where by comparing form factors it is possible
to obtain mesonic light front wave functions. In the present case 
it is not necessary to reinterpretate the
holographical coordinate $z$ as in standard LFH, where $z$ 
is the distance between constituent partons. 

The nucleon GPDs obtained have an exponential form, as in several 
phenomenological approaches, and their detailed form is typical  
for the limit of $x \to 0$.

In the future we plan to extend the formalism of AdS/QCD to obtain
other parton distribution functions of nucleons and also of other 
baryons, which could then be used in the evaluation of 
different hadronic processes. Note there exist in literature 
(see Refs.~\cite{Pire:2008zf,Marquet:2010sf}) 
preliminary results for deep inelastic scattering (DIS) and 
deeply virtual Compton scattering (DVCS) in AdS/QCD in case of scalar field. 
The authors have doubts on the applicability of the AdS/QCD 
framework to DIS and DVCS reactions and thus on the consistency 
of the AdS/QCD approach to access to GPDs. In particular, 
they stressed that one should try: 1) to include nonmininal coupling 
of string mode dual to observable hadron and electromagnetic bulk-to-boundary 
propagator; 2) to go beyond canonical dimension of the 
hadron operator (it means one can take into account its anomalous 
dimension). In additional, we would like to check other possibilities:  
to include nonconformal warping of the AdS metric which leads to modification 
of the effective dilaton potential of the soft-wall model and 
to test different dilaton profiles. 

\begin{acknowledgments}

The authors thank Tanja Branz, Stan Brodsky and Guy de T\'eramond 
for useful discussions.
This work was supported by FONDECYT (Chile) under Grants No. 3100028 
and No. 1100287, by the DFG under Contract No. FA67/31-2 and No. GRK683. 
This research is also part of the European Community-Research Infrastructure 
Integrating Activity ``Study of Strongly Interacting Matter'' 
(HadronPhysics2, Grant Agree\-ment No. 227431), Ru\-ssi\-an President grant 
``Scientific Schools'' No. 3400.2010.2, Federal Targeted Program 
``Scientific and sci\-en\-ti\-fic-pe\-da\-go\-gi\-cal 
personnel of innovative Russia''  Contract No. 02.740.11.0238. 

\end{acknowledgments}

\end{document}